\documentstyle[12pt]{article}

\title{Boundaries and junctions in two parity violating models in 2+1
dimensions}
\author{}
\date{\today}

\begin{document}
\bibliographystyle{unsrt}
\maketitle

\centerline{Mark Burgess}

\vspace{0.5cm}

\centerline{\em Oslo College of Engineering, Cort Adelersgata 30}
\centerline{\em N-0254 Oslo, Norway}

\centerline{and}

\centerline {\em Institute of Physics, University of Oslo}
\centerline {\em P.O. BOX 1048, Blindern, N-0316 OSLO 3, Norway}
\vspace{1cm}
\centerline{Margaret Carrington}
\vspace{0.5cm}
\centerline {\em Department of Physics and Winnipeg Institute for Theoretical
Physics}
\centerline{\em University of Winnipeg, Winnipeg, Manitoba}
\centerline{\em Canada R3B 2E9}

\begin{abstract}
Recently it has been suggested that junctions between materials with
different parity violating properties would be characterized by
diffusion layers, analogous to those in the p-n
junction\cite{burgess1,burgess2}.
This remark is amplified by a fuller investigation of two related
parity violating effective Lagrangians, which possess a kind of
duality. It is shown that gauge invariance and energy conservation
are sufficient to determine the behaviour at the interface.
This leads to modifications of normal parity-violating electrodynamics.
The coupling of an interface to an external system is a natural
solution to the deficiencies of Maxwell-Chern-Simons theory.
A heuristic model of a transistor-like device is discussed which
relates to recent experiments in device technology. Radiative
corrections to Chern-Simons theory induce a local magnetic moment
interaction whose lagrangian is everywhere gauge invariant. The
effects of this interaction are compared to Maxwell-Chern-Simons
theory. The dispersion of classical waves for these models is computed and
the laws of reflection and refraction are found to hold despite
the lack of $P$ and $T$ invariance. The magnetic moment dispersion is
gapless in contrast to the Chern-Simons dispersion except in the
case of a scalar
field which is covariantly constant. Both models
exhibit optical activity (Faraday effect).
\end{abstract}

\section{Introduction}
The Chern-Simons term is widely exploited in the
construction of effective theories
breaking parity and time-reversal invariance.
Although originally the term was introduced as a way of
providing the Yang-Mills gauge
field with a gauge invariant mass\cite{schonfeld1,deser1,deser2},
it has since appeared more often in
the condensed matter literature in discussions of systems like
the fractional quantum Hall effect and anyon superconductivity\cite{fradkin1}.
These models postulate the Chern-Simons term as an effective action for
an unknown microscopic theory, the coefficient of which
takes on a constant value which is chosen or, in principle, determined
from the underlying physics.

In this paper we explore more carefully the possible roles
of the Chern-Simons Lagrangian by investigating
systems in which its coefficient -- the physical potency and
sign of the
parity violating effects -- takes on different values
in different parts of the system. An additional model
which couples a gauge invariant current to the dual
of the field strength is also considered and compared to
the usual Chern-Simons term. This model reduces to the
Chern-Simons expression in a special case.

The junction scenario described
in this paper, although motivated
on theoretical grounds, could have some bearing on
recent experiments in which participating
electrons are characterized by predominantly a single
spin direction\cite{device1,device2}. Redlich has shown that
spin polarized Dirac fermions give rise to an effective Chern-Simons
theory when quantum corrections are accounted for\cite{redlich1} and
thus a Chern-Simons theory describes such a two-dimensional electron gas.

We begin by formulating the simplest
junction in terms of an action principle.
This consists of two regions in a (2+1)-dimensional
space, which meet on the line $x_1 = 0$. The two regions are
considered to have physically disparate properties so that
a boundary condition is implied for the physical fields
in addition to the relevant field equations on either side of the
boundary.

\section{Formalism}
To illustrate the variational formalism we are using,
consider the simplest case of plain electromagnetism at a material junction.
In Maxwell theory, the only variables in a physical medium
are the dielectric properties of matter; it is sufficient to
describe these properties in terms of a
conserved electric current variable
$J_{TOT}^\mu(x)$ whose value is position dependent
and a dipole field $P_{\mu\nu}\equiv D_{\mu\nu}-F_{\mu\nu}$. The action is
given by
\begin{equation}
S = \int dV_x \lbrace -\frac{1}{4}D^{\mu\nu}F_{\mu\nu}
-J_{TOT}^\mu A_\mu\rbrace\label{eq:1}
\end{equation}
where $F_{\mu\nu}=\partial_\mu A_\nu-\partial_\nu A_\mu$ ($F_1^{~2}=B$,
$F^{i0}=E^i$),
$dV_x = dtdx_1dx_2$ and the metric signature is
$\eta_{\mu\nu}=(+--)$. Units are chosen such that $\hbar=c=\epsilon_0=\mu_0=1$
and latin
indices refer to spatial dimensions only.
We shall have occasion to consider both sharp and soft interfaces.
To
take account of the polarization of media, one must allow the possibility
that currents will run along the surface of a sharp interface at
$x_1=0$. One can, with a certain freedom, either include this effect
in $P^{\mu\nu}$ or introduce the current explicitly as a surface current.
In the following we include both, anticipating the work in later sections.
Since we are modelling the interface as an abrupt change in
the physical properties of our system, these currents will be proportional to
a delta-function in the $x_1$ direction. We write
\begin{equation}
J^\mu_{TOT}= J^\mu+j^\mu\,\delta(x_1)=J^\mu+J_s^\mu\label{eq:2}
\end{equation}
$j^1=0$ is expected from the geometry (this is not the case in the Chern-Simons
theory) and the total
current will be conserved $\partial_\mu J_{TOT}^\mu = 0$. Varying the action
and
integrating by parts leads to
\begin{eqnarray}
\delta S &=& \int dV_x \lbrace \delta A^\nu\,\partial^\mu (F_{\mu\nu}
+P_{\mu\nu})
-J^\mu \delta A_\mu \rbrace\nonumber\\
&+& \int dV_x \lbrace \partial^\mu\left[ -\delta
A^\nu\,(F_{\mu\nu}+P_{\mu\nu})\right] -\delta A^\nu j_\nu\,\delta(x_1)\rbrace =
0
\label{eq:3}
\end{eqnarray}
Requiring that $A_\mu \rightarrow 0$ as $|x_\mu|\rightarrow \infty$ leads
to
\begin{eqnarray}
\delta S &=& \int dV_x \lbrace \delta A^\nu\,\partial^\mu D_{\mu\nu}
-J^\mu \delta A_\mu \rbrace\nonumber\\
&+& \int dtdx_2\int_{-\epsilon}^{+\epsilon} dx_1 \lbrace \partial^1\left[
-\delta A^\nu\,D_{1\nu}\right] -\delta A^\nu j_\nu\,\delta(x_1)\rbrace =
0\label{eq:4}.
\end{eqnarray}
Moreover, if the regular field equations (in the absence of a
boundary) are to apply arbitrarily close to $x_1=0$, then
the boundary condition
\begin{equation}
\delta S_{boundary}=\lim_{\epsilon\rightarrow 0}\int
dtdx_2\int_{-\epsilon}^{+\epsilon} dx_1 \lbrace \partial^1\left[ -\delta
A^\nu\,(F_{1\nu}+P_{1\nu})\right] -\delta A^\nu j_\nu\,\delta(x_1)\rbrace =
0\label{eq:5}
\end{equation}
is implied. Here it is assumed that the components of
the vector potential $A^\mu$ are continuous across the boundary.
This assumption may be unnecessarily
restrictive if one considers the case in
which a contact potential characterizes the junction
and will be relaxed later. This leads to the form
\begin{equation}
\Delta (F_{1\mu}+P_{1\mu})+j_\mu = 0 ~~~~~~\mu = 0,2\label{eq:6}
\end{equation}
where $\Delta$ means the change in value across the boundary. The
field equations outside of the boundary are given by
\begin{equation}
\partial^\mu D_{\mu\nu} = J_\nu\label{eq:7}
\end{equation}
Clearly one can absorb the contribution due to surface
currents into the more usual form of a polarization tensor.
The surface polarization tensor may be defined by
\begin{equation}
\partial^\mu P_{\mu\nu}^s = J_{\nu}^s\label{eq:8}
\end{equation}
which is added to $P_{\mu\nu}$.
The antisymmetric components of the
total polarization are $P_1^{~2} = -M$, $P^{10}=P^1$, enabling
(\ref{eq:6}) to be written in the standard form
\begin{eqnarray}
\Delta D^1 = \Delta(E^1+P^1) = 0\nonumber\\
\Delta H = \Delta(B-M) = 0\label{eq:9}.
\end{eqnarray}
where $P^1$ and $M$ are the polarization and magnetization respectively.
It also follows immediately from the assumption about the continuity of
the vector potential that
$\Delta E_2=\Delta (\partial_2 A_0 - \partial_0 A_2) = 0$.
The conservation of the total current gives us a relation between
$j_\mu$ and the change in $J_\mu$ across the boundary ($\Delta J_\mu$).
\begin{equation}
\partial^\mu (J_\mu + j_\mu \delta(x_1)) = 0\label{eq:11}
\end{equation}
Integrating this equation with respect to $x_1$ from $-\epsilon$ to
$+\epsilon$ and consider the limit $\epsilon\rightarrow 0$,
one obtains
\begin{equation}
\Delta J_1 + \partial^\mu j_\mu = 0,  ~~~~~~~~~(j_1 = 0)\label{eq:12}
\end{equation}

The main point to note from this exercise is that
it is unnecessary to explicitly introduce fields $D$ and $H$ to account
for polarization and magnetization effects specifically on the boundary:
it is sufficient to include the possibility of surface currents.
We shall therefore not refer to these fields again.

\section{Maxwell-Chern-Simons Theory}

\subsection{Gauge invariance at a junction}

The action formalism does not yield an explicitly
gauge invariant boundary condition when the Chern-Simons term is
considered. The Abelian Chern-Simons term is gauge invariant
only in the absence of boundaries, being quadratic
in $A_\mu$ but only linear in the derivatives.
Consider the action,
\begin{equation}
S = \int dV_x \lbrace -\frac{1}{4}F^{\mu\nu}F_{\mu\nu}
+\frac{\mu}{2}\epsilon^{\mu\nu\lambda}A_\mu \partial_\nu A_\lambda
-J_{TOT}^\mu A_\mu \rbrace\label{eq:13}
\end{equation}
The variation of this action, without further stipulation, leads to
the field equations and associated boundary condition
\begin{eqnarray}
\partial^\mu F_{\mu\nu}+\mu \epsilon_{\nu\rho\lambda}\partial^\rho A^\lambda
&=& J_\nu\label{eq:14}\\
\Delta \lbrack -F_{1\lambda} + \frac{1}{2}\mu\epsilon_{\mu 1
\lambda}A^\mu\rbrack
- j_\lambda  &=& 0\label{eq:15}
\end{eqnarray}
Gauge invariance of the boundary condition implies a restriction
on $A_\mu$. One is therefore led to consider the
effect of the gauge transformation $A'_\mu = A_\mu -\partial_\mu \theta$
on the action (\ref{eq:13}) over a region $-\epsilon < x_1 < \epsilon$ as
$\epsilon \rightarrow 0$.
\begin{equation}
\delta S=S[A_\mu']-S[A_\mu] = \int dtdx_2\int_{-\epsilon}^{+\epsilon} dx_1
\lbrace -\partial^1\lbrack \frac{1}{2}\mu \epsilon_{\nu 1\lambda}
\partial^\nu A^\lambda + J_1\rbrack - (\partial^\mu j_\mu)\delta(x_1)\rbrace
\theta\label{eq:16}
\end{equation}
where $j_1 = 0$, $j_0, j_2$ are independent of $x_1$
and $E^i = F^{i0}$. If the gauge choice $\theta$ is to be
arbitrary one has
\begin{equation}
\Delta (\frac{1}{2}\mu E_2 + J_1) + \partial^\mu j_\mu =0.\label{eq:17}
\end{equation}
On use of (\ref{eq:12}) this simply becomes
\begin{equation}
\Delta \mu \;E_2 = 0\label{eq:18}
\end{equation}
since the continuity of the vector potential implies that $\Delta E_2=0$.
This result indicates that the only valid boundary condition
between Chern-Simons media with different coefficients is one in which
$E_2=0$ on the boundary. This boundary condition
does not permit the passage of electromagnetic waves or information.
The vanishing of $E_2$ implies the form
\begin{equation}
A_\mu = \partial_\mu \xi(x) ~~~~~~~~~\mu\not=1\label{eq:19}
\end{equation}
for some scalar field $\xi(x)$. This solution may be used in the
action as a restriction on the allowed variations. The analogue of equation
(\ref{eq:5}) is then
\begin{equation}
\delta S_{boundary} = \int dt\,dx_2\int_{-\epsilon}^{+\epsilon} dx_1
\lbrace \partial^1\partial^\nu F_{1\nu}-\frac{1}{2}\partial_1(\mu
\epsilon_{\mu 1 \lambda} \partial^\lambda A^\mu) +\partial^\mu
j_\mu \delta(x_1)\rbrace \delta \xi=0\label{eq:20}
\end{equation}
which, on use of the current conservation equation, gives the gauge
invariant boundary condition
\begin{equation}
\Delta( \partial^\mu F_{\mu 1} +\frac{1}{4}\mu \epsilon_{\mu 1 \lambda}
F^{\lambda\mu}-J_1) =0\label{eq:21}
\end{equation}
This is seen to be consistent with the expression obtained from the field
equations (\ref{eq:14}), integrated directly over an infinitesimal region.

\subsection{Completing the action}

The gauge non-invariance of the Chern-Simons action at non-reflecting
junctions is an indication of the incompleteness of the Chern-Simons
theory at the boundary. Physically, in the region of a boundary
one expects short
wavelength modifications to come into play which will either modify the
values of physical fields or modify the action itself.
To rectify the omission one must either return to a more
fundamental theory and rederive the correct effective description, or
postulate the remainder of the degrees
of freedom in such a way that gauge invariance
is restored. A general requirement of the latter is the introduction of
new variables. We now wish to discuss this latter problem in more detail.
The general solution has been discussed by
one of us in\cite{burgess1,burgess2}, ignoring contact
potentials. Consider the action
\begin{equation}
S=\int dV_x \Bigg\lbrace -\frac{1}{4}F^{\mu\nu}{F_{\mu\nu}
+\frac{1}{2}\mu(x)\epsilon_{\nu\rho\lambda}A^\nu\partial^\rho A^\lambda + f(x)
J^\mu A_\mu}\Bigg\rbrace.\label{eq:22}
\end{equation}
Gauge invariance implies the restriction
\begin{equation}
\frac{\delta S}{\delta\theta}=\int dV_x\Bigg\lbrace
-\frac{1}{2}(\partial^\nu\mu)\epsilon_{\nu\rho\lambda}\partial^\rho A^\lambda
 - (\partial_\nu f)J^\nu\Bigg\rbrace = 0\label{eq:23}
\end{equation}
which may be written
\begin{equation}
\frac{1}{2}(\partial_t\mu)B + \frac{1}{2}(\partial_1\mu)E^2
=(\partial_t f)J^0+(\partial_1 f)J^1.\label{eq:24}
\end{equation}
The conserved external source $J_\mu$ can, if necessary, be
used to obtain a solvable
equation. If $\mu(x)$ is a function of say $x_1$ and $t$ then there are
sufficient terms to find a solution for $\mu$ in terms of $E^2$ and
$B$ without introducing the source.
The solution by this method will in the general case result
in the introduction of higher derivative terms in the action.

The role
of $f(x)$ is to act as a mediating `leaky membrane' which separates
the external
source from the two dimensional system. Thus although $J_\mu$ is
conserved in total, it may appear to be non-conserved via its
contact with the two dimensional junction.
Since the sole function of this
term is to balance the gauge invariance equation, a natural boundary
condition is the vanishing of $f(x)$ when $\partial_\nu\mu(x)=0$.
Taking the variation to be with respect to $x_1$ and
$t$ only, this is satisfied by
\begin{equation}
f(x_1,t) = \frac{1}{2}\alpha(\partial_1\mu)
+\frac{1}{2}\beta(\partial_t\mu)\label{eq:25}
\end{equation}
for constants $\alpha$ and $\beta$, hence
\begin{eqnarray}
\Big(
\alpha J^1 \partial_1^2 -E^2\partial_1
+\beta J^0\partial_t^2 -B\partial_t
\Big) \mu(x_1,t) +
(\alpha J^0+\beta J^1)\partial_1\partial_t\mu(x_1,t)=0.\label{eq:26}
\end{eqnarray}
This equation is too difficult to solve in the most general case,
but some insight can be gained by noting that space and time
appear symmetrically. Considering time-independent solutions,
representing steady state solutions one has
\begin{equation}
\alpha J^1 (\partial_1^2\mu)-E^2(\partial_1\mu)=0.\label{eq:27}
\end{equation}
Letting $y = \partial_1\mu$ and $P(x_1)=E^2/\alpha J^1(x_1)$, one has
\begin{equation}
\frac{dy}{dx^1}-P(x^1)y=0\label{eq:28}
\end{equation}
which can be cast as a total differential by introducing the
integrating factor $\exp(-\int^{x^1}_0 P(z)dz)$,
\begin{equation}
\frac{d}{dx^1}\Bigg( y e^{-\int^{x^1}_0 P(z)dz}\Bigg)=0\label{eq:29}
\end{equation}
thus one has
\begin{equation}
\frac{d\mu}{dx^1} = \frac{\mu_c}{L} e^{\int^{x^1}_0 P(z) dz} \equiv
\frac{d}{dx^1}\Bigg(
g(x^1)e^{\int^{x^1}_0 P(z) dz}
\Bigg)\label{eq:30}
\end{equation}
where $\mu_c$ and $L$ are constants.
Comparison of these equations leads to a differential equation for $g(x^1)$,
\begin{equation}
\frac{dg(x^1)}{dx^1}+g(x^1)P(x^1) = \frac{\mu_c}{L}.\label{eq:31}
\end{equation}
If, not unreasonably, $E^2$ and $J^1$ are approximately correlated, then one
may write $P(x)=p+\Delta P(x)$, for constant $p$ and the equation for $g(x^1)$
takes the form
\begin{equation}
\frac{dg(x^1)}{dx}+pg(x^1)=\frac{\mu_c}{L} - \Delta P(x^1)g(x^1)\label{eq:32}
\end{equation}
whose general solution is given by
\begin{equation}
g(x^1) = \int_0^{x^1} dz e^{p(z-x^1)}(\frac{\mu_c}{L}- \Delta P(z)g(z))
+g_0 e^{-px^1}.\label{eq:33}
\end{equation}
When $\Delta P(x^1)=0$, this is simply the exponential decay law. In general,
the exponential behaviour is modulated by a self-consistently defined
function $\Delta P$, but for any `physical' function $\Delta P$, the
long term behaviour will always be dominated by the exponential factors.
Further assumptions of symmetry can be made in order to simplify even
more\cite{burgess2}. The key observation is that the diffusion-like
equation (\ref{eq:27}) implies that the Chern-Simons parameter must fall off
exponentially in space. Clearly, from the symmetry of (\ref{eq:26}), the same
argument
also applies for the space-independent equation for time-variation.
The field equations in the varying region must now be solved self-consistently
so as to record the effect of the constraint introduced through the solution
of the gauge-invariance condition.

Had the boundary between the regions been curved rather
than linear then it is natural to
suppose that the decay law would be modified by the extrinsic curvature of
the interface\cite{burgess5}. Since the concentration of field is greater
around a
`corner' like feature, this would lead to an intensification of $E^1$
in this region. Although this would not seem to modify the invariance
constraint it will, though the field equations, exacerbate the current $J^1$
in the vicinity of the interface. This increased activity must be answered by
the sources, thus one would expect increased dissipation in such as region.

\subsection{Examples}
A simple example of the foregoing procedure can be computed in the absence of
sources\cite{burgess2}. When $\mu=\mu(x_1,t)$ and $E_2$ and $B$ are assumed
to be constants throughout the region of interest, the gauge invariance
constraint may be solved together with the Bianchi identity to show that
both the Chern-Simons coefficient and $E_1$ are arbitrary functions of
$\gamma(E_2^{-1}x_1-B^{-1}t)$, for constant $\gamma$.
This solution is extremely general and
admits both longitudinal waves and exponential decay, but does not
correspond to any obvious physical situation. One possibility is that it
is a heuristic representation of periodic impurities in a quantum Hall system.

An example of more physical relevance is the case of a time-varying
Chern-Simons coefficient in a time-varying magnetic field.
Since it is known that
a system of spin-polarized fermions gives rise to an effective field
theory involving a Chern-Simons term\cite{redlich1}
in the long wavelength limit,
this situation should correspond directly to spin relaxation, or spin pumping
in a 2+1 dimensional system. Also,
the construction of a model of two coupled systems in which spin
migrates from one half-plane to the other could be composed of
two systems of this kind, with a third junction layer of space- and
time-varying
$\mu$ describing the contact region.
The constraint that $\int d^2x \mu(x)$ be conserved
for all times is a natural addition,
implying that what leaves one half-plane must end up
in the other. This corresponds to the conservation of spin in the
spin picture. Of course this is
only one physical interpretation of such a system: the effective field
equations know nothing of any microscopic origins and are therefore
not prejudiced by the association with spin or any other parity violating
effect.

To solve the formal constraint, let
$\mu=\mu(B(t),t,\rho)$, $\rho = J^0$, where $J^\mu$ is the source
in (\ref{eq:22}). The constraints on the variables in the action are
Bianchi identity
\begin{equation}
\partial_t B + \epsilon_{ij}\partial_i E^j=0\label{bt:1}
\end{equation}
and the gauge invariance identity (\ref{eq:26})
\begin{equation}
\frac{\delta S}{\delta\theta} = \int dV_x\Bigg\lbrace\beta \rho \ddot{\mu} -
B\dot{\mu}\Bigg\rbrace=0
\label{bt:2}
\end{equation}
where dots represent partial time derivatives and
the boundary condition (\ref{eq:25}) implies that
\begin{equation}
f(t) = \frac{1}{2}\beta\dot{\mu}.\label{bt:3}
\end{equation}
Eqn. (\ref{bt:2}) may be solved for $\mu$ as in the preceeding
subsection. It is important to note that the role of
the gauge invariance identity is two-fold here.
It is both an algebraic relation between variables in the action and
a condition on the possible variations of the action. Since $J_\mu$ is
an external source, one has that $\delta J_\mu=0$, thus a variation of
the action leads to
\begin{eqnarray}
\delta S&=&\int dV_x\Bigg\lbrace
\delta A^\nu \partial^\mu F_{\mu\nu}+\frac{1}{2}\frac{\delta\mu}{\delta
B}\delta B
\epsilon^{\mu\nu\lambda} A_\mu\partial_\nu A_\lambda
+\mu\epsilon^{\mu\nu\lambda}\delta A_\mu \partial_\nu A_\lambda\nonumber\\
&+&\frac{1}{2}(\partial_\nu\mu)\epsilon^{\mu\nu\lambda}\delta A_\mu A_\lambda
+\frac{\delta f(t)}{\delta B}\delta B(J^\mu A_\mu)+f(t)J^\mu \delta A_\mu
\Bigg\rbrace\label{bt:4},
\end{eqnarray}
where $\delta B = \epsilon_{ij}\partial_i \delta A^j$.
The variation of the constraint (\ref{bt:2}) yields
\begin{equation}
\Bigg( \beta\rho\frac{\delta\ddot{\mu}}{\delta B}
-\dot{\mu}-B\frac{\delta\dot{\mu}}{\delta B}\Bigg) \delta B= 0\label{bt:5}.
\end{equation}
The only apparent solution
is $\delta B=0$, which implies that the vector potential may be written
$\delta A^j = \partial^j\xi$ for some scalar function $\xi$ and $j\not=0$.
The zeroth component $A^0$ is unrestricted and hence there is no
contradiction with (\ref{bt:1}). Substituting this into the variation
of the action (\ref{bt:4}) gives the field equations for the system.
For $A^0$ one has,
\begin{equation}
\frac{\delta S}{\delta A^0} = \int dV_x \Bigg\lbrace \partial_i E^i - \mu(B,t)
B(t) + f(t)\rho \Bigg\rbrace= 0
\label{bt:6}
\end{equation}
which is identical in form to the usual result for constant $\mu$. For $A^j$
one has
\begin{eqnarray}
\delta S &=& \int dV_x \Bigg\lbrace
(\partial^i\xi)(\partial^\mu F_{\mu i})+\mu\epsilon^{i\nu\lambda}
(\partial_i\xi)\partial_\nu A_\lambda\nonumber\\
&-&\frac{1}{2}\dot{\mu}\epsilon^{ij}(\partial_i\xi)A_j
+f(t)J^i(\partial_i\xi)
\Bigg\rbrace =0\label{bt:7}
\end{eqnarray}
Integrating by parts and using the conservation equation
$\partial^i J_i=-\dot\rho$ gives
\begin{equation}
B + \beta \dot{\rho}=0\label{bt:8}
\end{equation}
which agrees precisely with the gauge invariance condition (\ref{bt:2}) up to
a total time-derivative. Thus the
consistency between the
field equation and the gauge invariance
condition is restored, by analogy with (\ref{eq:21}). It is interesting
to observe that the field equation is independent of the Chern-Simons
coefficient, so that linearity is preserved. The equation (\ref{bt:2})
determining the coefficient takes the form
\begin{equation}
\dot{\mu} = C \exp \int \frac{B(t)}{\beta\rho} dt\label{bt:9}
\end{equation}
which exhibits exponential behaviour.
The connection with spin relaxation can now be noted as follows. The
variable $B$ in the action is the effective electromagnetic field, not
the microscopic field felt by the spins. This includes the effect of the spin
degrees of freedom. Since the coefficient of the Chern-Simons term is
proportional in some sense to the sign of the spin and depends on the chemical
potential of the spins\cite{burgess1,burgess10} this is the relevant variable
to consider. One is thus interested in consistent solutions of (\ref{bt:9}).
As the effective field $B$ tends to zero, the Chern-Simons
coefficient tends to a constant value (typically zero). For changing
$B$ and $\mu$ the external source is needed to drive the
system. The simplest solution is for exponentially decaying $\mu$ and
$B$ which corresponds to spin relaxation in the absence of an
external field.
If one drives the spin system with an adiabatically sinusoidal
time-varying magnetic field,
this is reflected by an oscillatory part
for $B$. This is coupled to the time-variation of $\mu$ through
(\ref{bt:6}) and leads to an exponential lag in the response,
corresponding to a hysteresis effect.

\subsection{Contact potential: a switching junction}

A further example of physical interest arises in the case of a
sharp boundary supporting a contact potential $\Delta A_0$. Taking the
action (\ref{eq:13}) and boundary condition (\ref{eq:15}) one has,
in components
\begin{eqnarray}
\Delta(B-\frac{1}{2}\mu A_0)-j^2=0\nonumber\\
\Delta(-E_1+\frac{1}{2}\mu A_2)-j^0=0\label{eq:34}
\end{eqnarray}
and there is a step $\Delta\mu$ at $x_1=0$. This system possesses certain
qualities resembling those of a transistor or switching device: two
regions of differing properties separated by a thin junction to
which an external (bias) current is applied. It is
straightforward to show that the picture
has the properties of a switching device. Taking $j^\mu$ to be an
external source (coupling to the third dimension, for instance) which acts only
at the junction, conventional electromagnetic
boundary conditions are obtained (and gauge invariance is restored)
provided
\begin{eqnarray}
\frac{1}{2}\Delta\mu \overline A_0 + \overline \mu \Delta A_0 &=&
-j^2\label{eq:35}\\
\frac{1}{2}\Delta\mu \overline A_2 &=& j^0\label{eq:36}
\end{eqnarray}
where barred quantities signify the mean values of the respective
parameters at the discontinuity ($\overline A_2=A_2$).
Let $\phi\equiv\Delta A_0$. Since
the step in the Chern-Simons coefficient is a physical quantity,
it should be gauge invariant, thus equating
(\ref{eq:35}) and (\ref{eq:36}) one has
\begin{equation}
-\frac{j^0}{\overline A_2} = \frac{j^2 + \overline\mu\phi}{\overline
A_0}\label{eq:37}
\end{equation}
and for gauge invariance
\begin{equation}
\frac{\overline A_0}{\overline A_2} = \frac{\overline
A_0+\partial_0\theta}{\overline A_2+\partial_2\theta}\label{eq:38}.
\end{equation}
This implies a restriction on the gauge invariance of the theory
at the boundary to
transformations of the form\footnote{For instance, $\theta$ might
be of the form $\exp(i(kx^2+\omega t))$ in which case it plays the
role of a massless excitation.}
\begin{equation}
\theta = \theta(\overline A_2 x^2+\overline A_0 t)\label{eq:39}.
\end{equation}
This condition is satisfied by wavelike solutions, for instance. The
gauge invariant ratio
$\alpha\equiv \overline A_0 / \overline A_2$ must be
regarded as a property of a
given interface and the gauge invariance requirement becomes
\begin{equation}
\phi = -\frac{j^2+j^0\alpha}{\overline \mu}.\label{eq:40}
\end{equation}
The contact potential or $\Delta\mu$ is seen to depend on the external current
$j^2$
as well as the density at the boundary $j^0$. In particular, in the absence of
current, the step must collapse.

Now suppose that an external electrostatic potential $V$ is held across
this junction in reverse bias (opposing the potential $\phi$). Donor
charges (i.e. those not taken into account by the effective theory) or
quasi-particles
will only be able to surmount the potential barrier only if $V > \phi$.
However the size of $\phi$ is controlled by the external source
and thus the junction can be made to switch a small electric current. It is
noted, on the other hand, that this `transistor' is somewhat
unusual since the apparent magnetic field leads also to a Hall drift of the
electrons, and thus they drift parallel to the interface as well as
across it.
The generic behaviour described here is clearly relevant in a device which is
populated largely by electrons of the same spin, or in an ordinary
device with differing magnetizations in a strong magnetic field.
It is known that, in a system of fermions
with predominantly one spin direction,
a Chern-Simons term is induced by radiative corrections\cite{redlich1}
in the long wavelength limit, the coefficient of which is proportional
to the two-dimensional spin eigenvalue. The above scenario then describes
a spin-switch, i.e. a semi-conductor-like switch which is controlled
entirely by parity-violating spin effects. Some experimental evidence for
this exists already.

The flow in and out of the system by sources at the boundary, in
our notation, has the appearance of a charged current. This is because
$J_\mu$ is formally conserved. However, the variable coupling $f(x)$
implies that $J_\mu$ is not conserved on the boundary. There, the
lack of manifest gauge invariance implies that the compensating current
could appear neutral, as seen from the junction's perspective.
Thus $J_\mu$ need not be interpreted as a current of electrons.
The dissipation could relate to radiated energy
generated by spin flip transitions in moving from one region to
the other. If an electric current is prevented from flowing
freely then it would be expected, from the preceding
discussion, that there would be some
resistance to the transport across the junction, since
a tendency toward a reflective
boundary condition would arise. This could lead to a component of
$J_1$ arising, implying through (\ref{eq:26}) an accumulation of
spins at the junction. While these would decay by diffusion eventually,
there is the possibility of some hysteresis depending on the conductivity
of the material at the boundary.

Finally, it is important to bear in mind that the present theoretical
model exists only at the level of an effective field theory, not a microscopic
one.
The motion of dressed quasi-particles through an interface of changing
physical properties must take place through the interaction with
some third party, since the particles must be `undressed' and `redressed'
in order to make
the transition across the barrier. In the above construction, this occurs
through an interaction with the source.
The source itself consists formally of
quasi-particles characterized by the mediating value of the physical
parameter $\overline\mu$.
In the above discussion we have adopted the classical viewpoint of an
effective field theory
to avoid dealing with the subtleties of quantum tunneling, but in the final
analysis one is interested in dispensing with the effective theory and
finding the appropriate microscopic one. Tunneling processes will be
relevant at this level.

\subsection{Experimental evidence}

After predicting the diffusive junction behaviour and the above switch model,
we learned of two separate experiments which have a direct relevance to these
findings. Experiments by Kane et al.\cite{device1} consider the interface
between two quantum Hall systems with different filling fractions\footnote{We
thank B. Halperin and T. Finstad for pointing out these references.}.
This corresponds to differing Chern-Simons coefficients according to the
arguments in refs. \cite{burgess1,zhang1}. The key observation is that
the aligned spins break parity invariance and that the magnitude of this
breaking depends on the number of spins\cite{burgess10}.
Here one observes
a diode-like behaviour, as expected from the general arguments above. It is
interesting that this experiment shows evidence for a polarization of
nuclear spins at the junction. This coupling to nuclear
spin acts as source (sink) of the polarization in crossing the junction
and constitutes the `external system' in the preceding source
language\cite{burgess1,burgess2}.

A second set of experiments by Johnson relates more directly to the switching
phenomenon\cite{device2}.
Here a device is constructed in which spin-polarized
charge carriers inhabit a thin gold layer with a two dimensional symmetry.
Spins subsequently migrate across a junction into a region of different or
opposite spin\cite{device2} by passing through a finite width junction which is
coupled
to an external source of current. A
`spin bottleneck' phenomenon is observed which prevents the
non-equilibrium junction from decaying when the source is active
and switching is indeed observed, in
accordance with the above discussion.

\section{Non-minimal coupling}

In this section we study a related system which avoids some of the gauge
invariance problems of the Chern-Simons system.

\subsection{Gauge Invariance}
Consider the action
\begin{equation}
S = \int dV_x \Bigg\lbrace -\frac{1}{4}F^{\mu\nu}F_{\mu\nu}
+\frac{1}{2}\kappa(x)
\epsilon^{\mu\nu\lambda}J_\mu F_{\nu\lambda}\Bigg\rbrace\label{eq:41}
\end{equation}
for some conserved current $J^\mu$. This model has been studied in a number
of recent works\cite{stern1,kogan1,carrington1,carrington2,carrington3}.
The coupling
of the current to the dual of the field strength leads to a local
magnetic moment
type interaction for scalar particles\cite{kogan1}.
In 2+1 dimensions this leads to an induced
phase analogous to the anyon phase, known as the Aharonov-Casher
phase\cite{carrington3}. An interesting duality exists between this term and
the usual Chern-Simons type parity breaking term. If one begins with scalar
electrodynamics coupled to a Chern-Simons term, the non-minimal coupling
is induced by radiative corrections\cite{kogan1}.
The coefficient is related to the inverse of the Chern-Simons
coefficient and the electric charge and does not vanish in the
long wavelength limit.
Similarly, if one
begins with the non-minimal coupling and computes one-loop corrections,
a regular Chern-Simons term is induced. Apart from these
properties, this non-minimal coupling is interesting in the present
context due to its manifest gauge invariance, even in the presence
of a boundary. It is natural to compare the boundary properties of the above
model with the more usual Chern-Simons term.

Applying the action principle to (\ref{eq:41}), one has
\begin{equation}
\partial_\mu \lbrack F^{\mu\lambda} +\kappa(x) \epsilon^{\mu\nu\lambda}
J_\nu\rbrack = 0\label{eq:42}
\end{equation}
and associated boundary condition, for a junction at $x_1=0$,
\begin{equation}
\Delta(F^{1\lambda}+\kappa(x)\epsilon^{1\nu\lambda}J_\nu) =0.\label{eq:43}
\end{equation}
It is now possible to compare the action (\ref{eq:41}) to the Chern-Simons
action coupled to a source (\ref{eq:22}). A direct comparison is only possible
if the current itself includes terms involving the gauge field. This
is the case for a complex scalar field, for example. Then one has
\begin{equation}
J_\mu = i \lbrack\Phi(D_\mu\Phi)^\dagger
-\Phi^\dagger(D_\mu\Phi)\rbrack,\label{eq:44}
\end{equation}
where $D_\mu=\partial_\mu-ieA_\mu$,
or in the unitary parameterization $\Phi=\rho e^{i\theta}$,
\begin{equation}
J_\mu = 2\rho^2\lbrack (\partial_\mu\theta)-eA_\mu\rbrack.\label{eq:45}
\end{equation}
The non-minimal term in (\ref{eq:41}) then takes the form
\begin{equation}
2\kappa(x)\rho^2 F^{*\mu}(\partial_\mu\theta-eA_\mu)\label{eq:46}
\end{equation}
where $F^{*\mu}=\frac{1}{2}\epsilon^{\mu\nu\lambda}F_{\nu\lambda}$.
Comparing this to (\ref{eq:22})
and derived quantities, it is possible to identify
\begin{eqnarray}
\mu(x) &=& -4e\kappa(x)\rho^2\label{eq:47}\\
f(x)J^\lambda &=&
-2\epsilon^{\mu\nu\lambda}(\partial_\nu\kappa\rho)(\partial_\mu\theta)\label{eq:48}
\end{eqnarray}
which indicates that the non-minimal coupling can be regarded as
a normal Chern-Simons term with a variable coefficient
plus an external massless scalar field
$\theta$ which lives in the region $\nabla \mu(x)\not=0$ (for example
on boundaries). This is consistent with the discussion in section 3.2 and in
refs. \cite{burgess1,burgess2}.

\subsection{Atomic spring model}

For the purpose of determining bound state properties for comparison
with ref. \cite{burgess3},
it is interesting to consider a toy model of a two dimensional
medium in which electrons are bound to their parent atoms by means
of a pseudo-harmonic potential. If $s^i$ is the displacement vector
of a single electron then under a Lorentz force,
\begin{equation}
(\partial_t^2+\gamma\partial_t+\omega_0^2)s^i=-\frac{e}{m}(E^i+\epsilon_{ij}B_c\partial_t s^j)\label{eq:49}
\end{equation}
where $B_c$ is a constant external magnetic field, $\omega_0^2=k/m$ for
spring constant $k$ and $\gamma$ is a damping factor. The current
arising from this motion can be characterized by
\begin{equation}J^\mu =\Bigg(
\begin{array}{c}
\rho  \\
-Ne\partial_t \vec{s}
\end{array}\Bigg)\label{eq:50}
\end{equation}
where $\rho$ is the total charge density (which is usually zero, including
the effect of background charges) and $N$ is
the number of optically active electrons. Adding a minimal gauge coupling
$J^\mu A_\mu$ to (\ref{eq:41}), and omitting dielectric
polarization effects one has the field equation for constant $\kappa$
\begin{equation}
\partial_\mu F^{\mu\lambda}+\kappa\epsilon^{\mu\nu\lambda}\partial_\mu J_\nu
=J^\lambda\label{eq:51}
\end{equation}
and boundary condition given by (\ref{eq:43}). We shall use this model below
in order to determine the optical properties of this system. Such properties
were previously calculated for the usual Chern-Simons term in ref.
\cite{burgess3}.

\subsection{Complex Scalar Field}
A more realistic quantizable field theory can be obtained by considering
a scalar field coupled minimally to a gauge field. This is closely
related to the effective $P$ and $T$ breaking theory introduced in ref.
\cite{wen2}
for superconductivity,
and the to the Landau-Ginsburg theory of the Hall effect\cite{zhang1}. The
scalar
field represents the collective excitations of an unknown microscopic
system in the long wavelength limit.
In the Chern-Simons case, this model has already been
shown to lead to a Faraday rotation in reflected waves\cite{wen2}.
The principal difference from the
atomic spring model is that the scalar field is a superconductor-insulator.
There is no in-built mechanism for dissipation at zero temperature, hence no
finite conductivity.  Moreover, the model does not describe bound states.
In the superconductive regime, waves will not
propagate inside the material, but extra-planar waves can be reflected
off a two-dimensional surface in which the scalar field lives. In the
insulating regime, wavelike-solutions can only exist under special
circumstances which will be described below.

The action for a complex scalar field is defined by
\begin{equation}
S = \int dV_x \Bigg\lbrace(D^\mu\Phi)^\dagger(D_\mu\Phi)-m^2\Phi^\dagger\Phi
-\frac{\lambda}{6}(\Phi^\dagger\Phi)^2
-\frac{1}{4}F^{\mu\nu}F_{\mu\nu}+\frac{1}{2}\kappa
\epsilon^{\mu\nu\lambda}J_\mu F_{\nu\lambda}   \Bigg\rbrace\label{eq:52}
\end{equation}
where the current is given by (\ref{eq:44}) and $D_\mu = \partial_\mu-ieA_\mu$.
It is noteworthy that the matter coupling involves a
quadratic dependence on the gauge field that cannot
be written strictly in the form $J^\mu A_\mu$. The
field equation for the scalar field is given by
\begin{equation}
(D^2+m^2)\Phi + 2i\kappa
F^{*\mu}(D_\mu\Phi)+\frac{\lambda}{3}(\Phi^\dagger\Phi)\Phi
+i(\partial^\mu\kappa)F^*_\mu\Phi=0\label{eq:53}
\end{equation}
with associated boundary condition at $x_1=0$,
\begin{equation}
\Delta(D_1+\frac{i}{2}\kappa\epsilon_{1\nu\lambda}F^{\nu\lambda})\Phi=0.\label{eq:54}
\end{equation}
For the gauge field, one obtains
\begin{equation}
\partial_\mu F^{\mu\lambda}+\kappa\epsilon^{\mu\nu\lambda}\partial_\mu J_\nu
-2e\kappa F^{*\lambda} \Phi^\dagger\Phi=e J^\lambda\label{eq:55}
\end{equation}
provided
\begin{equation}
\Delta(F^{1\lambda}+\kappa\epsilon^{1\nu\lambda}J_\nu) = 0\label{eq:56}.
\end{equation}

\section{Energy Momentum Tensor}

The gauge non-invariance of the Chern-Simons Lagrangian at a boundary
is accompanied by a discontinuity in the electromagnetic Poynting
vector $\Delta S_1=\Delta(E_2 B)$\cite{burgess1,burgess2}. This is
another indication that the theory is not complete at the boundary.
To properly understand the energy flow in the present models one
should supplement the usual electromagnetic Poynting flow by a
contribution from the parity violating term. A consideration of
the energy-momentum tensor as defined through N\" other's theorem
leads to the relevant conserved quantities\footnote{If one defines
the energy-momentum tensor by $T^{\mu\nu}=\frac{\delta S}{\delta g_{\mu\nu}}$
for the metric $g_{\mu\nu}$ then there is no contribution from
parity violating terms, since $\epsilon^{\mu\nu\lambda}$ transforms like a
tensor density and is subsequently independent of the metric. However,
this should be regarded as a failure of the variational definition
rather than a reason to disregard the extra terms.}. The generalized
force law for the system is given by
\begin{equation}
f^\mu =\int dv_x\partial_\nu\theta^{\nu\mu}\label{emt:1}
\end{equation}
which, in the case of the electromagnetic field, gives rise to the
Lorentz force law $f^i = -\int dv_x J^\mu F_{~\mu}^{i}$, where
$dv_x$ is a spatial volume element.
Any anomalies in $\theta^{\mu\nu}$ could give rise to corrections to
this force law and are therefore important to the discussion of boundary
effects. A generic feature of the energy-momentum tensor in the
present models is that it is non-symmetric and gauge-dependent. This
signals a breakdown of Lorentz invariance as well as gauge invariance.
That these two principles should be broken simultaneously is reminiscent
of the deficiencies of the canonical energy momentum
tensor\cite{jackiw2,eriksen1} and thus in what follows we shall adopt the
covariant
procedure of refs. \cite{jackiw2} and \cite{eriksen1} to obtain directly
the normally symmetric `Belifante energy-momentum tensor'.

To construct $\theta^{\mu\nu}$, one notes that the covariant vector
potential $A^\mu$ transforms like a 3-vector only up to a gauge
transformation. N\" other's theorem implies
that under a Lorentz transformation
\begin{eqnarray}
x^\mu\rightarrow {x^\mu}'&=& x^\mu + \delta x^\mu\label{emt:2}\\
A^\mu(x) \rightarrow {A^\mu}'(x) &=& A^\mu(x)+\delta A^\mu(x)\label{emt:3}
\end{eqnarray}
a symmetric theory is characterized by the continuity equation $\partial_\mu
C^\mu=0$, where
\begin{equation}
C^\mu = \frac{\partial {\cal L}}{\partial (\partial_\mu A_{\sigma})}\delta
A_\sigma + {\cal L}\delta x^\mu\label{emt:4}
\end{equation}
where $S = \int dV_x {\cal L}$. The defining equation for $\theta^{\mu\nu}$
is then
\begin{equation}
C^\mu = -\theta^{\mu\nu}\delta x_\nu.\label{emt:5}
\end{equation}
The gauge-Lorentz invariant restriction
$\delta A_\mu=F_\mu^{~\sigma}\delta x_\sigma$ implies that
\begin{equation}
\theta^{\mu\nu} =  -\frac{\partial {\cal L}}{\partial
(\partial_\mu A_{\sigma})} F_{\sigma}^{\,\,\,\nu}-\eta^{\mu\nu}{\cal
L}\label{emt:6}.
\end{equation}
If a boundary is introduced then Lorentz invariance is
explicitly violated and this expression loses its immediate interpretation
as a consequence of the lack of translational invariance. In the
absence of a parity violating term one has the usual electromagnetic
energy momentum tensor
\begin{equation}
\theta^{\mu\nu}_{EM} = F^{\sigma\mu}F^\nu_{~\sigma} +
\frac{1}{4}\eta^{\mu\nu}F^{\rho\sigma}F_{\rho\sigma}.\label{emt:7}
\end{equation}
The corrections $\Delta\theta^{\mu\nu}$ to this due to
the parity violating terms will now be discussed.

\subsection{Chern-Simons}

Use of the Chern-Simons Lagrangian in (\ref{emt:6}) gives
\begin{equation}
\Delta\theta^{\mu\nu} = \frac{1}{2}\mu \epsilon^{\rho\mu\sigma} A_\rho
F^\nu_{~\sigma}
-\frac{1}{4}\mu\eta^{\mu\nu}\epsilon^{\rho\sigma\lambda}A_\rho
F_{\sigma\lambda}.
\label{emt:8}\end{equation}
This contribution is not symmetric, but, using the Bianchi
identity $\partial_{(\mu}F_{\nu\lambda)}=\epsilon^{\mu\nu\lambda}\partial_\mu
F_{\nu\lambda}=0$ it is seen to be gauge-invariant provided
the coefficient $\mu$ is a constant. The integral of the divergence
of the full $\theta^{\mu\nu}=\theta_{EM}^{\mu\nu}+\Delta\theta^{\mu\nu}$ is
conserved in the absence of sources,
\begin{equation}
\int dV_x \partial_\mu\theta^{\mu\nu} = 0\label{emt:9}
\end{equation}
where the field equations (\ref{eq:14}) have been used. Adding sources
produces the conventional modification
\begin{equation}
\int dV_x \partial_\mu\theta^{\mu\nu} = -\int dV_x J^\sigma
F^\nu_{~\sigma}\label{emt:10}
\end{equation}
which is the standard Lorentz force law. The latter result implies that
no modification of the force law is required in Maxwell-Chern-Simons
theory, except at boundaries or in regions of variable $\mu(x)$ in
which case one finds that
\begin{equation}
\int dV_x \partial_\mu\theta^{\mu\nu} = \int dV_x \Bigg\lbrace
\mu(x) F^{*\sigma}(\frac{1}{2}F^\nu_{~\sigma}+\frac{1}{2}(\partial^\nu
A_\sigma))
\Bigg\rbrace.\label{emt:11}
\end{equation}
Under a gauge transformation $A_\mu\rightarrow A_\mu+\partial_\mu\chi$, this
term changes by the
gradient of $-\frac{1}{2}(\partial^\sigma\mu(x))F^*_\sigma\chi$,
which may be compared to (\ref{eq:23}) for zero source. This identification
shows explicitly that the lack of energy conservation at the
boundary is
related to the lack of gauge invariance, as previously argued in
ref. \cite{burgess1} and leads to a generalized force on the system.

In ref. \cite{jackiw1} the presence of an instability in the dispersion was
found. Although some evidence for this can be seen in $\theta^{00}$ (which
has no definite sign), this does not cause any problems in the present work.

\subsection{Non-minimal term}

For the non-minimal Lagrangian, one has
\begin{equation}
\Delta\theta^{\mu\nu}=\kappa\epsilon^{\rho\mu\sigma}J_\rho F^\nu_{~\sigma}
-\frac{1}{2}\eta^{\mu\nu}\kappa\epsilon^{\alpha\beta\lambda}J_\alpha
F_{\beta\lambda},
\label{emt:12}
\end{equation}
and it is assumed that $J_\mu$ is independent of $A_{\mu,\nu}$.
This correction is non-symmetric, but is explicitly gauge invariant.
The integral over the divergence of the total energy-momentum tensor
is conserved, provided $\kappa$ is
constant, and the addition of sources leads to the usual Lorentz
modification (\ref{emt:10}). For non-constant $\kappa(x)$, one has
\begin{equation}
\int dV_x \partial_\mu\theta^{\mu\nu} = -\int dV_x \kappa
\epsilon^{\rho\mu\sigma}\Bigg\lbrace (\partial_\mu J_\rho)
F^\nu_{~\sigma}-\frac{1}{2}(\partial^\nu
J_\rho)F_{\mu\sigma}\Bigg\rbrace\label{emt:13}.
\end{equation}
This term is gauge invariant and represents a modification of
the standard Lorentz force wherever $\kappa$ is a function of $x^\mu$.

In ref. \cite{burgess1} it was shown that a Casimir force is necessarily
present on a gauge-invariant reflective boundary due to quantum or thermal
fluctuations in Maxwell-Chern-Simons theory. The force arises because there
is a mass or gap-mismatch in the spectra for the two sides of the junction.
Here the spectrum of the non-minimal
coupling is notably gapless (massless) in the absence of an ordered phase
(spontaneous symmetry breaking). See equation (\ref{eq:86}). This seems also
to be confirmed by refs. \cite{carrington1,carrington2} at one loop, but here
one notes that scalar and gauge loops are inextricably linked by the
non-minimal term so that it is not possible to consider the gauge field
in isolation. Further investigations would be required to determine the absence
of a gap due to fluctuations in general.

Finally, $\theta^{00}$ does not have a definite sign, which suggests the
possibility of an instability for certain values of the sources. No instability
is found in the dispersion (\ref{eq:86},\ref{eq:100}) however. In the case
of normally impinging waves for inhomogeneous currents,
this is less clear (\ref{eq:103}).

\section{Electromagnetic waves}

In an earlier paper\cite{burgess3} one of us has considered
the properties of wavelike
solutions of the Lagrangian (\ref{eq:13}). Related work has also been
carried out in $3+1$ dimensions\cite{jackiw1}.
There it was remarked that a
future problem would be to consider the analogue of Fresnel's equations
at a material interface. Since the naive boundary condition for
the Chern-Simons model is not gauge invariant, it is not immediately
clear how to proceed. To satisfy the gauge invariant boundary
condition (\ref{eq:18}), there must be total reflection from the
line $x_1=0$. If on the other hand one couples to an external source as
in (\ref{eq:22}), then the gauge non-invariant parts of the boundary condition
vanish
and one is left with the normal electromagnetic boundary conditions. The
virtue of the non-minimal model is its automatic gauge invariance at the
boundary.

\subsection{Maxwell-Chern-Simons atomic spring model}

We begin with the unmodified Chern-Simons theory and consider the case in
which the coefficient is constant.
\subsubsection{Planar dispersion relation}
The dispersion relation is obtained on substituting the trial solution
\begin{eqnarray}
E^i&=&E^i_0e^{i(k_ix_i-\omega t)}\label{eq:57}\\
B&=&B_0e^{i(k_ix_i-\omega t)}+B_c\label{eq:58}\\
J^i&=&J^i_0e^{i(k_ix_i-\omega t)}\label{eq:59}\\
\rho &=& \rho_0e^{i(k_ix_i-\omega t)}+\rho_c\label{eq:60}
\end{eqnarray}
into the components of the field equations
\begin{eqnarray}
B_c&=&-\frac{1}{\mu}\rho_c\label{eq:61}\\
ikE_{0\|}&=&\mu B_0+\rho_0\label{eq:62}\\
i\omega E_{0\|}-\mu E_{0\perp}&=&J_{0\|}\label{eq:63}\\
-ik B_0+i\omega E_{0\perp}+\mu E_{0\|} &=& J_{0\perp}\label{eq:64}
\end{eqnarray}
where the parallel and perpendicular projections are defined through
the relations
\begin{eqnarray}
k^iE_0^i&=&kE_{0\|}\nonumber\\
\epsilon_{ij}k^iE_0^j&=&kE_{0\perp}\label{eq:65}
\end{eqnarray}
and similar ones for the current. If one neglects the oscillatory
part of the magnetic field from the equations of motion (which implies
that the magnetically induced current-response is small) then these
equations, together with (57) and (58), can be manipulated so as to eliminate
all variables except the electric field, which then satisfies the equation
\begin{equation}
\left(
\begin{array}{cc}
i\omega mW+iNe^2\omega -i\mu\omega eB_c & -\mu mW -eB_c k^2+\omega^2eB_c\\
mW-\omega^2 eB_c & (\frac{-ik^2}{\omega}+i\omega)mW +iNe^2\omega -i\mu\omega
eB_c
\end{array}
\right)
\left(
\begin{array}{c}
E_{0\|} \\
E_{0\perp}
\end{array}
\right) = 0\label{eq:66}
\end{equation}
where $W = (-\omega^2-i\gamma\omega+\omega_0^2)$. Demanding
the vanishing of the determinant of this matrix leads immediately
to the dispersion relation
\begin{equation}
k^2=\omega^2\Bigg\lbrace 1-\frac{\mu^2}{\omega^2}
+\omega_p^2\frac{W(1+\frac{\mu^2}{\omega^2})+\omega_p^2-2\mu\omega_L}
{W^2+\omega_p^2W-\omega^2\omega_L^2}\Bigg\rbrace\label{eq:67}
\end{equation}
where $\omega_p^2 = Ne^2/m$ and $\omega_L = eB_c/m$. This result has
been given in ref. \cite{burgess3}. The refractive index is given
by $n=k/\omega$.  The real and imaginary parts of the refractive index are
plotted against $\omega$ for various values of $\mu$ in Figs. 1a and 1b.
At a junction between
two regions, this dispersion
relation applies provided the interface is sharp. Sources are then
needed to balance the requirements of gauge invariance, but their
introduction leads only to normal electromagnetic boundary conditions,
thus there is no modification to the laws of reflection or refraction
and Fresnel's relations are given by the usual formulas (see below).
Thus when two dimensional waves strike the interface, currents are
set up along the interface. These currents must either disperse into
the two dimensional system or
out into the external system. If no
sources are introduced, the dispersion is only given by (\ref{eq:67}) at $t=0$.
The subsequent decay of the interface then modifies the dispersion
in a non-linear way until a situation of equilibrium is reached.

\subsubsection{Extra-planar dispersion relation}

To show that the Chern-Simons term gives rise to a modified
Faraday effect\footnote{This is also referred to as optical activity
or circular birefringence.}, one embeds the planar model into a
three dimensional space and directs plane polarized waves so that
they impinge normally to the plane. Since the matter fields describe an
ostensibly two dimensional system
all currents are restricted to the plane. The embedded action is given by
\begin{equation}
S=\int dV_{\hat{x}} \Bigg\lbrace -\frac{1}{4}F^{\hat{\mu}\hat{\nu}}
F_{\hat{\mu}\hat{\nu}} -J^{\mu} A_{\mu}
+\frac{1}{4}\tilde{\mu}\epsilon_{\hat{\mu}\hat{\nu}\hat{\lambda}3}
A^{\hat{\mu}}F^{\hat{\nu}\hat{\lambda}} \Bigg\rbrace\label{eq:68}
\end{equation}
where careted indices run over the third spatial dimension in addition
to their usual values and $\tilde{\mu}$ has dimension one.
The components of the field equation
are then given by
\begin{eqnarray}
\partial_{\hat{i}}E^{\hat{i}}-\tilde{\mu} B^3 &=&\rho\nonumber\\
-\partial_t E^{\hat{i}}+\epsilon_{\hat{i}\hat{j}\hat{k}}\partial_{\hat{j}}
B^{\hat{k}}
-\tilde{\mu}\epsilon_{\hat{i}\hat{j}3}E^{\hat{j}} &=&J^i\nonumber\\
\partial_t B^{\hat{k}}+\epsilon_{\hat{i}\hat{j}\hat{k}}\partial_{\hat{i}}
E^{\hat{j}}&=&0\nonumber\\
\partial_{\hat{i}}B^{\hat{i}}&=&0.\label{eq:69}
\end{eqnarray}
Next consider a solution of the form
\begin{equation}
E^{\hat{i}}=E_0^{\hat{i}}e^{i(kz-\omega t)}\label{eq:70}
\end{equation}
with corresponding expressions for the magnetic field and current.
To obtain the dispersion relation between $k$ and $\omega$ it is
sufficient to consider the planar components of the electric field.
It is convenient to define complex variables $E = E^1+jE^2$ where
$j^2=-1$, but $ij =ji\not=-1$\footnote{Note that if these
imaginary units are treated as anticommuting numbers they reproduce
the su(2) algebra of rotations. Here, as commuting numbers they
represent a realization of the group $U(1)\times U(1)$.} and corresponding
variables for the displacement vector $s$\cite{burgess4}. With these variables
the
equation of motion for the electrons (\ref{eq:49}) becomes (neglecting damping
terms)
\begin{equation}
(-\omega^2+\omega_0^2)s = -\frac{e}{m}(E-jB_c\partial_t s)\label{eq:71}.
\end{equation}
Combining the second and third equations in (\ref{eq:69})
gives
\begin{equation}
Ne\omega^2 s=(\omega^2-k^2)E -ij\omega\mu E.\label{eq:72}
\end{equation}
Finally, eliminating one of the variables from the last two equations
gives the dispersion relation
\begin{equation}
k^2 = \omega^2\Bigg\lbrace
1+\omega_p^2\frac{-\omega^2+\omega_0^2}{(-\omega^2+\omega_0^2)^2-\omega_L^2\omega^2}
-ij(\frac{\tilde{\mu}}{\omega}+\omega_p^2\frac{\omega\omega_L}{(-\omega^2+\omega_0^2)^2-\omega_L^2\omega^2})
\Bigg\rbrace.\label{eq:73}
\end{equation}
The real and imaginary parts of the refractive index $n=k/\omega$ are plotted
against $\omega$ for various values of $\tilde{\mu}$ in Figs. 2a and 2b.
The combination $ij$ guarantees energy conservation and implies that the
wavenumber has the form
\begin{equation}
k = k_r -ij k_{ij}\label{eq:74}
\end{equation}
which shows that the transmitted wave suffers a rotation of
its plane of polarization
\begin{equation}
E = E_0e^{i(k_rz-\omega t)}(\cos \,k_{ij}z +j \sin\,k_{ij}z ).\label{eq:75}
\end{equation}
It is normal to define the  measurable rotation angle in terms of
`Verdet's constant' $V$ by
\begin{equation}
\chi_{rot} = VB_cL\label{eq:76}
\end{equation}
where $L$ is the distance travelled through a stack of planar systems,
but since the Chern-Simons term results in such a rotation even in
the absence of $B_c$, it is sufficient to consider the angle rotated
per unit length $k_{ij}$.

\subsection{Non-minimal atomic spring model}

\subsubsection{Planar dispersion relation}
Wavelike solutions to the non-minimal action (\ref{eq:41}) are obtained
by the same method as in ref. \cite{burgess3}. It is curious to
note that the presence of the curl of the current makes the field
equations sensitive to anisotropy and inhomogeneities in the the
matter field. The components of (\ref{eq:51}) give,
\begin{eqnarray}
\partial_i E^i-\kappa\epsilon_{ij}\partial_i J^j=\rho\label{eq:77}\\
-\partial_t E^i + \epsilon_{ij}\partial_j B+\kappa\epsilon_{ij}
(\partial_j\rho+\partial_tJ^j)=J^i\label{eq:78}
\end{eqnarray}
Using the wave ansatz (\ref{eq:57})-(\ref{eq:60}), and considering only the
oscillating parts leads to
\begin{eqnarray}
ik E_{0\|}-i\kappa k J_{0\perp}&=&\rho_0\label{eq:80}\\
i\omega E_{0\|}-i\omega\kappa J_{0\perp}&=&J_{0\|}\label{eq:81}\\
i\omega E_{0\perp}-ikB_0-i\kappa k\rho_0+i\kappa\omega
J_{0\|}&=&J_{0\perp}\label{eq:82}\\
ikE_{0\perp}=i\omega B_0\label{eq:83}
\end{eqnarray}
where the last equation follows from the (\ref{bt:1}).
The continuity equation for the current
(which is not an independent equation, but
follows from (\ref{eq:80}) and (\ref{eq:81}) above) may be written
\begin{equation}
kJ_{\|0}=\omega\rho_0.\label{eq:84}
\end{equation}
Eliminating the electric field in favour of the current, one
obtains after some effort the matrix equation
\begin{equation}
\left(
\begin{array}{cc}
mW+Ne^2 & iNe^2\omega\kappa-i\omega eB_c\\
i\omega e B_c-iNe^2\omega\kappa & mW
+Ne^2\frac{\omega^2}{\omega^2-k^2}
\end{array}
\right)
\left(
\begin{array}{c}
J_{0\|}\\
J_{0\perp}
\end{array}
\right) = 0\label{eq:85}
\end{equation}
The vanishing of the determinant of this matrix yields the dispersion
relation
\begin{equation}
k^2 = \omega^2\Bigg\lbrack 1+\frac{\omega_p^2
W+\omega_p^4}{W^2+\omega_p^2W-\omega^2(\omega_L-\kappa\omega_p^2)^2}\Bigg\rbrack\label{eq:86}.
\end{equation}
The real and imaginary parts of the refractive index $n=k/\omega$ are plotted
against $\omega$ for various values of $\kappa$ in Figs. 3a and 3b.

It is noteworthy that, on making the
definitions
\begin{eqnarray}
D^i &=& E^i-\kappa\epsilon_{ij}J^j\nonumber\\
H &=& B+\kappa\rho\label{eq:87}
\end{eqnarray}
the wave equation for the transverse magnetic waves takes the usual
form
\begin{equation}
(\partial_t^2-\nabla^2)H=\epsilon_{ij}\partial_i J^j,\label{eq:88}
\end{equation}
and the Bianchi identity is unchanged, implying that the normal laws
of reflection and refraction apply:
\begin{eqnarray}
\frac{\sin\theta_t}{\sin\theta_i} &=& \frac{n_1}{n_2}\label{eq:89}\\
\theta_i &=& \theta_r\label{eq:90}
\end{eqnarray}
where the angles $\theta$ refer to the incident, transmitted and
reflected angles to the normal. It is quite possible that a non-linear
scalar field theory (for instance, with a $\lambda\phi^4$ term or higher power)
would violate this law, if wave solutions can even by found.

{}From the linearity, it is possible to write
\begin{equation}
(E_{0i}-E_{0r})\cos\theta_i =E_{0t}\cos\theta_t\label{eq:91}
\end{equation}
Thus, defining the wave impedance $Z=E/H=\mu_r/n^2$, where
$\mu_r H=B$, and
noting that the boundary condition may be written $\Delta H=0$, one
has the familiar result that
\begin{equation}
\frac{E_{0i}}{Z_1}+\frac{E_{0r}}{Z_1}=\frac{E_{0t}}{Z_2}.\label{eq:92}
\end{equation}
Combining (\ref{eq:91}) and (\ref{eq:92}) gives Fresnels standard relations for
the transmission and reflection coefficients
\begin{eqnarray}
\frac{E_{0r}}{E_{0i}}=\frac{Z_1cos\theta_i-Z_2\cos\theta_t}{Z_2\cos\theta_t+Z_1\cos\theta_i}\label{eq:93}\\
\frac{E_{0t}}{E_{0i}}=\frac{2Z_2\cos\theta_i}{Z_2\cos\theta_t+Z_1\cos\theta_i}.\label{eq:94}
\end{eqnarray}

\subsubsection{Extra-planar dispersion relation}
To compute the Faraday effect, the planar theory is embedded in a
three dimensional space, with
waves impinging normally, as before. The action is therefore
\begin{equation}
S=\int dV_x \Bigg\lbrace -\frac{1}{4}F^{\hat{\mu}\hat{\nu}}
F_{\hat{\mu}\hat{\nu}} -J^{\mu} A_{\mu}
+\frac{1}{2}\tilde{\kappa}\epsilon_{\hat{\mu}\hat{\nu}\hat{\lambda}3}
J^{\mu}F^{\hat{\nu}\hat{\lambda}} \Bigg\rbrace\label{eq:95}
\end{equation}
where careted indices include the third dimension. The currents are assumed
to lie purely in the plane. The components of the field equation and
Bianchi identity
are now given by
\begin{eqnarray}
\partial_{\hat{i}}E^{\hat{i}}-\tilde{\kappa}\epsilon_{ij}\partial_i J^j&=&\rho
\nonumber\\
-\partial_t E^{\hat{i}}+\epsilon_{\hat{i}\hat{j}\hat{k}}\partial_{\hat{j}}
B^{\hat{k}}+\tilde{\kappa}\epsilon_{ij}(\partial_j \rho+\partial_t J^i)&=&J^i
\nonumber\\
\partial_t B^{\hat{k}}+\epsilon_{\hat{i}\hat{j}\hat{k}}\partial_{\hat{i}}
E^{\hat{j}}&=&0\nonumber\\
\partial_{\hat{i}}B^{\hat{i}}&=&0.\label{eq:96}
\end{eqnarray}
The oscillating magnetic field can be eliminated to yield a wave equation
for the electric field
\begin{equation}
E^{\hat{i}}=E_0^{\hat{i}} e^{i(kz-\omega t)}.\label{eq:97}
\end{equation}
given by
\begin{equation}
(-\partial^2_t+\nabla^2)E^{\hat{i}}+\tilde{\kappa}\epsilon_{ij}(\partial_t
\partial_j \rho+\partial_t^2 J^j)=\partial_t J^i.\label{eq:98}
\end{equation}
It is pertinent to note that the presence of spatial derivatives of the
current in the above expression implies that the system is sensitive
to inhomgeneities and anisotropy. A proper description of anisotropy
cannot be obtained from the foregoing equations, since the spring
constant $k=\omega_0^2 m$ would need to be different in the $x_1$
and $x_2$ directions to make the assumption self-consistent.

\subsubsection{Homogeneous isotropic medium}

In a homogenous, isotropic medium, the spatial derivatives of the
current vanish identically.
Using the complex coordinate representation introduced earlier
 one has
\begin{equation}
(\omega^2-k^2) E +ij\tilde{\kappa}\omega^3 Nes  = \omega^2 Nes\label{eq:99}.
\end{equation}
Using the equation of motion for $s$ (\ref{eq:71}), it
is possible to eliminate $E$ giving immediately the dispersion relation
\begin{equation}
k^2=\omega^2\Bigg\lbrack 1+\frac{\omega_p^2(1-ij\tilde{\kappa}\omega)}
{(-\omega^2+ij\omega_L\omega+\omega_0^2)}\Bigg\rbrack.\label{eq:100}
\end{equation}
or
\begin{eqnarray}
k^2 =\omega^2\Bigg\lbrace
1 + \frac{\omega_p^2\lbrack \omega_0^2-\omega^2+\tilde{\kappa}\omega_L\omega^2
-ij(\tilde{\kappa}\omega(\omega_0^2-\omega^2)+\omega_L\omega)
\rbrack}{(\omega_0^2-\omega^2)^2-\omega^2\omega_L^2}
\Bigg\rbrace
\label{eq:101}
\end{eqnarray}
The real and imaginary parts of the refractive index $n=k/\omega$ are plotted
against $\omega$ for various values of $\tilde{\kappa}$ in Figs. 4a and 4b.

$k_{ij}$ is the rotation per unit length through a layered system,
where $k = k_r - ijk_{ij}$.

\subsubsection{Homogeneous, anisotropic medium}

If the wavelength of waves is small compared to the inhomogeneities of the
material medium, the dispersion becomes sensitized to the structure.
Consider the case of a uni-axial crystal which permits inhomogenities
of the current in a preferred direction $x_1$. Using the trial
solution $E=E_0\exp(i(k_zz+k_1x_1-\omega t))$
one easily obtains
\begin{eqnarray}
(\omega^2-k_z^2-k_1^2) E^1-iNe\kappa\omega^3 s^2&=&Ne\omega^2 s^1\nonumber\\
(\omega^2-k_z^2-k_1^2) E^2 +i\omega \tilde{\kappa} Ne(\omega^2-k_1^2)s^1
&=& Ne\omega^2s^2.\label{eq:102}
\end{eqnarray}
The complex method is not appropriate here, owing to the
lack of symmetry. Nevertheless, it is possible to eliminate the
electric field component-wise and solve the determinental equation
for the matrix coefficient of $s^1$ and $s^2$, giving
\begin{equation}
k_z^2 = \omega^2 \Bigg\lbrace
1 - \frac{k_1^2}{\omega^2} +\frac{b\mp\sqrt{b^2-4ac}}{2a\omega^2}
\Bigg\rbrace
\label{eq:103}
\end{equation}
where $a = \omega_0^2-\omega^2-\omega^2\omega_L\tilde{\kappa}$,
$b = -2\omega_p^2\omega^2((\omega_0^2-\omega^2)+\tilde{\kappa}\omega_L
(\frac{k_1^2}{2\omega^2}-\omega) )$
and $c=\omega_p^4\omega^4(1+\tilde{\kappa}^2(k_1^2-\omega^2))$. The dispersion
continues to exhibit birefringence, but now with a marked asymmetry.
The $k_1$ terms either reduce or enhance the electron mobility, depending
on the sign of the magnetic field $\omega_L$. In particular, it is seen
that $k_1$ acts as an effective mass gap for the dispersion.

\subsubsection{Ohmic conductors}

The conducting limit of the previous results could in principle be
obtained from the $\omega^2_0\rightarrow 0$ limit of the atomic
spring model. It is useful to reexpress the result in terms of the
more familiar conductivity. Let the anisotropic conductivity tensor
be defined by its projected components
\begin{eqnarray}
J_{\perp} &=& \sigma_\perp E_\perp\nonumber\\
J_\|&=&\sigma_\|E_\|,\label{eq:104}
\end{eqnarray}
then from (\ref{eq:80})-(\ref{eq:83}) one has
\begin{equation}
\left(
\begin{array}{cc}
1+\frac{i}{\omega}\sigma_\| & -\kappa\sigma_\perp \\
\kappa\sigma_\| & 1+\frac{i\omega\sigma_\perp}{\omega^2-k^2}
\end{array}
\right)
\left(
\begin{array}{c}
E_{0\|} \\
E_{0\perp}
\end{array}
\right) = 0\label{eq:105}
\end{equation}
giving the dispersion relation
\begin{equation}
k^2=\omega^2\Bigg\lbrack
1+\frac{-\kappa^2\sigma^2_\|\sigma^2_\perp +i(\sigma_\perp\omega
+\sigma_\|^2\sigma_\perp/\omega+\kappa^2\omega\sigma_\|\sigma_\perp^2)}
{\omega^2(1+\kappa^2\sigma_\|\sigma_\perp)^2+\sigma^2_\|}
\Bigg\rbrack\label{eq:106}
\end{equation}
which reduces to standard results on setting $\kappa=0$, $\sigma_\|=0$.
What is interesting here is that the longitudinal current plays a role
in the dispersion. In the vicinity of a boundary, like the edge
of a finite sample, the simple split into $\sigma_\|$ and $\sigma_\perp$
must break down. Close to the edge, the longitudinal conductivity
must tend toward zero and be replaced by an enhanced transverse
conductivity. This corresponds to a modulation of the charge at the
edge of the sample, which then spreads out to form surface density waves.

The Faraday effect is straightforwardly obtained from (\ref{eq:99})
on substituting
$J=\sigma E$; we ignore the role of anisotropy and inhomogeneities here.
Then, straightforwardly
\begin{equation}
k^2 = \omega^2\Bigg\lbrack 1+ j\kappa\sigma +i\frac{\sigma}{\omega}\Bigg\rbrack
\label{eq:107}
\end{equation}
which indicates that the rotation of the polarization plane is accompanied
by damping and reflection. It is interesting to note that the
combination $j\kappa$ implies that the reflective
properties are unaffected by $\kappa=0$ whereas the polarization effect is
entirely due to $\kappa\not=0$.

\subsection{Non-minimal Complex scalar field}

Wavelike solutions for the electromagnetic field
are not a general feature of the
non-minimally coupled scalar field theory. Consider
the case in which the collective field mode is covariantly constant
i.e. $D_\mu\Phi=0$. Then the scalar field equation implies that
\begin{equation}
\Phi^2 = -\frac{3}{\lambda}(i\partial^\mu\kappa\;F^*_\mu+m^2).\label{eq:108}
\end{equation}
If the field strength is oscillating, this makes most sense
when $\partial_\mu\kappa=0$. In a superconducting phase there is clearly
no wave propagation in 2 dimensions. There is a regime however in which
propagating solutions can be obtained.
For covariantly constant $\Phi$
the current vanishes but $\Phi^\dagger\Phi$ is an invariant constant.
The field equations for the electromagnetic field are then
\begin{equation}
\partial_\mu F^{\mu\lambda} - 2e\kappa F^{*\lambda}\Phi^\dagger\Phi=0.
\end{equation}
Taking the spatial components of this equation together with the Bianchi
identity (\ref{bt:1}) leads to a determinental equation for the
dispersion relation
\begin{equation}
k^2=\omega^2 - 4e^2\kappa^2(\Phi^\dagger\Phi)^2
\end{equation}
which is notably
similar to that for the superconductor model in ref. \cite{wen2}, indeed
it is regular Chern-Simons dispersion for $\mu = 2e\kappa\Phi^\dagger\Phi$.

\section{Discussion}
We have examined
some of the consequences of parity violation in the vicinity of junctions
and boundaries. Using the principles of gauge invariance and energy
conservation, we derive the acceptable behaviour of two models for an
effective $P$ and $T$ breaking theory. In the case of regular
the Chern-Simons term, it is found that dissipation must be a feature
close to a boundary. This dissipation can either be a destructive
dissipation -- that is, one which tends to erode the boundary
itself, or a stable transfer of current to an external system by mediating
sources. It is interesting to note that, in the quantum Hall system,
edge currents can be measured and that dissipation is observed at the
boundaries of the sample, concentrated at the corners, which is at least
in qualitative agreement with the picture conveyed here. The non-minimal
coupling has by nature edge currents, but these are related unambiguously
to the magnetic moment coupling and result in no dissipation. On the
other hand, the presence of a `curl' of the current implies a mixing
of charge at the boundary with the bulk charge.

The diffusion-like behaviour of a regular Chern-Simons interface, together
with the need for an external source of current suggests a transistor
like behaviour. While this behaviour is in itself amusing, since it is derived
from a consistency argument in an
effective theory, it appears to have relevance to experimental devices in which
the charge carriers in different isolated regions have predominantly the same
spin\cite{device1,device2}. An electron passing from one region to another must
then flip spin, requiring either an input or a drain of energy.

We have considered the effect of penetration of
material samples by electromagentic radiation. Waves inside $P$ and $T$
breaking media are no longer transverse, but no essential modification
of the usual laws of reflection or refraction is noted. Dispersion
is qualitatively different for the two models considered. In the
regular Chern-Simons theory, the Chern-Simons parameter appears
mainly as a mass or gap term in the dispersion relation. The relation for
the non-minimal model does not appear to possess such a gap.
Both models exhibit the required optical activity, or circular birefringence.

Finally we note that similar studies of boundary effects in parity violating
models have been made in refs \cite{callan1,redlich2} and \cite{wen1}. In the
former
case it is shown that a connection exists between the chiral anomaly in
$2n+2$ dimensions and the Chern-Simons gauge anomaly in $2n+1$ dimensions, at
least to first order in a derivative expansion; when
chiral models are calculated non-pertubatively, one also finds that the
Chern-Simons
form is modified by higher order terms
and becomes the $\eta$-invariant. In the latter case, edge currents are argued
by appealing to linear response theory and gauge invariance. In both of these
cases, currents are responsible for balancing the gauge invariance constraints.
In this paper further solutions are found which do not require the inclusion
of additional currents and a new physical interpretation is given to the
gauge invariance problem.

\section*{Acknowledgement}

M.B. would like to thank P. Kelly, C. Korthals-Altes, T. Finstad, B. Halperin,
R. Jackiw
and L. Pryadko for pointing out a number of relevant reference.

\bibliographystyle{unsrt}

\vspace*{1.0cm}

\section*{Figure Captions}
\begin{description}
\item[1a.] The real part of the refractive index $n=k/\omega$ as determined
from Eqn. (88) plotted
against $\omega$ for various values of $\mu$.  We use $\omega_0=1$,
$\omega_p=.1 \omega_0$, $\omega_L=.01\omega_0$ and $\gamma=.01\omega_0$.\\
\item[1b.] The imaginary part of the refractive index $n=k/\omega$ as
determined
from Eqn. (88) plotted
against $\omega$ for various values of $\mu$.  We use $\omega_0=1$,
$\omega_p=.1 \omega_0$, $\omega_L=.01\omega_0$ and $\gamma=.01\omega_0$.\\
\item[2a.] The real part of the refractive index $n=k/\omega$ as determined
from Eqn. (94) plotted
against $\omega$ for various values of $\tilde{\mu}$.  We use $\omega_0=1$,
$\omega_p=.1 \omega_0$, $\omega_L=.01\omega_0$ and $\gamma=0$.\\
\item[2b.] The imaginary part of the refractive index $n=k/\omega$ as
determined
from Eqn. (94) plotted
against $\omega$ for various values of $\tilde{\mu}$.  We use $\omega_0=1$,
$\omega_p=.1 \omega_0$, $\omega_L=.01\omega_0$ and $\gamma=0$. \\
\item[3a.] The real part of the refractive index $n=k/\omega$ as determined
from Eqn. (106) plotted
against $\omega$ for various values of $\kappa$.  We use $\omega_0=1$,
$\omega_p=.1 \omega_0$, $\omega_L=.01\omega_0$ and $\gamma=.01\omega_0$.\\
\item[3b.] The imaginary part of the refractive index $n=k/\omega$ as
determined from Eqn. (106) plotted
against $\omega$ for various values of $\kappa$.  We use $\omega_0=1$,
$\omega_p=.1 \omega_0$, $\omega_L=.01\omega_0$ and $\gamma=.01\omega_0$.\\
\item[4a.] The real part of the refractive index $n=k/\omega$ as determined
from Eqn. (121) plotted
against $\omega$ for various values of $\tilde{\kappa}$.  We use $\omega_0=1$,
$\omega_p=.1 \omega_0$, $\omega_L=.01\omega_0$ and $\gamma=0$.\\
\item[4b.] The imaginary part of the refractive index $n=k/\omega$ as
determined
from Eqn. (121) plotted
against $\omega$ for various values of $\tilde{\kappa}$.  We use $\omega_0=1$,
$\omega_p=.1 \omega_0$, $\omega_L=.01\omega_0$ and $\gamma=.0$.
\end{description}
\end{document}